\PassOptionsToPackage{usenames,dvipsnames}{xcolor}
\documentclass{openjournal}
\pdfoutput=1 %for arXiv submission
\usepackage[T1]{fontenc}
\usepackage{ae,aecompl}
\usepackage[utf8]{inputenc}
\usepackage{newtxtext,newtxmath}
\usepackage[english]{babel}
\usepackage{amsmath,amstext,mathtools}
\usepackage{hyperref}
\usepackage[hang]{footmisc}
\usepackage{threeparttablex}
\usepackage{longtable}
\usepackage{inconsolata}
\usepackage{enumitem}
\usepackage{lineno}
\newcommand{\irresponse}[1]{{#1}}
\newcommand{\refresponse}[1]{#1}

\newcommand\blfootnote[1]{%
  \begingroup
  \renewcommand\thefootnote{}\footnote{#1}%
  \addtocounter{footnote}{1}%
  \endgroup
}

\shorttitle{PSFs of coadded images}
\shortauthors{Mandelbaum et al.\ (LSST~DESC)}

\begin{document}
\title{PSFs of coadded images}

\author{Rachel Mandelbaum\altaffilmark{1}, Mike Jarvis\altaffilmark{2}, Robert H. Lupton\altaffilmark{3}, James Bosch\altaffilmark{3}, Arun Kannawadi\altaffilmark{3}, Michael D.\ Murphy\altaffilmark{1}, Tianqing Zhang\altaffilmark{1}}
\noaffiliation

\collaboration{The LSST Dark Energy Science Collaboration}
\blfootnote{Author affiliations may be found before the references.}

%\affiliation{McWilliams Center for Cosmology, Department of Physics, Carnegie Mellon University, Pittsburgh, PA 15213, USA}
%
%\author{Mike Jarvis}
%\affiliation{Department of Physics \& Astronomy, University of Pennsylvania, 209 South 33rd Street, Philadelphia, PA 19104-6396, USA}
%
%\author{Robert H. Lupton}
%\affiliation{Princeton University, Princeton, NJ, USA}
%
%\author{James Bosch}
%\affiliation{Princeton University, Princeton, NJ, USA}
%
%\author{Arun Kannawadi}
%\affiliation{Princeton University, Princeton, NJ, USA}
%
%\author{Michael D.\ Murphy}
%\affiliation{McWilliams Center for Cosmology, Department of Physics, Carnegie Mellon University, Pittsburgh, PA 15213, USA}
%
%\author{Tianqing Zhang}
%\affiliation{McWilliams Center for Cosmology, Department of Physics, Carnegie Mellon University, Pittsburgh, PA 15213, USA}

%\author{Others to include?}

%\author{The LSST Dark Energy Science Collaboration}
%\noaffiliation

%\date{\today}

\begin{abstract}
    \noindent We provide a detailed exploration of the connection between choice of coaddition schemes and the \irresponse{point-spread function (PSF)} of the resulting coadded images.  In particular, we investigate what properties of the coaddition algorithm lead to the final coadded image having a well-defined PSF.  The key elements of this discussion are as follows: 
    \begin{enumerate}[leftmargin=2.0cm,rightmargin=1.3cm]
        \item We provide an illustration of how linear coaddition schemes can produce a coadd that lacks a well-defined PSF even for relatively simple scenarios and choices of weight functions.  
        \item We provide a more formal demonstration of the fact that a linear coadd only has a well-defined PSF in the case that either (a) each input image has the same PSF or (b) the coadd is produced with weights that are independent of the signal.
        \item We discuss some reasons that two plausible nonlinear coaddition algorithms (median and clipped-mean) fail to produce a consistent PSF profile for stars.
        \item We demonstrate that all nonlinear coaddition procedures fail to produce a well-defined PSF for extended objects.
    \end{enumerate}
    \irresponse{In the end, we conclude that, for any purpose where a well-defined PSF is desired, one should use a linear coaddition scheme with weights that do not correlate with the signal and are approximately uniform across typical objects of interest.}
\end{abstract}

\section{Introduction}\label{sec:intro}

For many purposes, \irresponse{astronomers} currently use and will in \irresponse{the} future use coadded images \irresponse{(i.e., images that combine the information from multiple single-epoch observed images by registering and stacking the images in some way)} to measure the properties of stars and galaxies observed in astronomical images.  The subject of this paper is the PSF of the coadded images, and in particular, {\em \irresponse{how does the coaddition algorithm determine whether or not the coadded image has a well-defined PSF, and how is that PSF defined?}}  We will henceforth use the term ``coadd'' to refer to the coadded image and refer to its PSF as the ``coadd PSF''.  \refresponse{In this context, ``well-defined'' PSF means that there is some function such that each location in the coadd can be described as the convolution of that function with the true sky scene (see text above Eq.~\ref{eq:coaddpsf} for details). } The answer to \refresponse{the italicized } question \refresponse{above} depends on (a) whether the coaddition scheme is linear or not (e.g., mean versus median), and (b) what weights are used for a linear coaddition scheme (e.g., including the object signal in an inverse variance weight). In particular, we will show that use of a nonlinear coaddition scheme and/or use of weights that include the signals of individual objects when carrying out a linear coaddition scheme results in a coadd that does not have a well-defined PSF. \irresponse{The authors were particularly motivated by the challenge of analyzing data from the Vera C.\ Rubin Observatory Legacy Survey of Space and Time \citep[LSST;][]{2009arXiv0912.0201L,2019ApJ...873..111I}, which will have hundreds of exposures at each point within the survey footprint, but the considerations described herein are applicable to any telescope and imaging survey.}

For this purpose, we will ignore other relevant issues that occur in practice, such as the need to obtain an accurate astrometric solution in each exposure before producing the coadd, issues that can arise when carrying out the image registration and resampling through interpolation\footnote{This problem is particularly non-trivial for undersampled images\irresponse{, such as those from space telescopes \citep[see discussion in][]{2011ApJ...741...46R}}.}, elimination of artifacts in individual exposures, magnitude-dependent detector effects that can modify the apparent PSF measured from stars, etc.  In other words, this paper assumes that (a) the individual images are perfectly aligned with each other, (b) the PSF and its spatial variation are perfectly known in each individual exposure contributing to the coadd, (c) all detector non-idealities have been perfectly corrected, (d) the coadd is produced at the native pixel scale\refresponse{; and (e) the sky background is known and perfectly uniform}. 
%Our formalism will, in addition, omit noise terms that are always present for real images. 
However, we will not assume that the individual exposures have the same PSF, \irresponse{since the} variation in \irresponse{the} PSFs between exposures is what makes this problem of producing a coadd with a meaningful and well-understood PSF challenging. \refresponse{Note that we make these assumptions not because we think they are realistic, but rather because we are trying to emphasize that even in a very simplified case, the coaddition process is non-trivial in complexity.  It is, however, well worth investigating, because  coadds can serve as a sufficient static for the static sky \citep{2017ApJ...836..188Z}.  Even coaddition algorithms that result in coadds that are not sufficient statistics may not lose much information in practice  \citep[e.g.,][]{2018PASJ...70S...5B}, and use of coadds can substantially reduce the computational costs of many common analyses (e.g., model fitting methods that require a model evaluation for each image in order to compute the likelihood -- one model evaluation per step in the fit to the coadd, and many per step in the fit for joint analysis of individual exposures).}
%Also, we will briefly touch on the issues that can arise when the coaddition algorithm is designed to eliminate artifacts by employing some nonlinear approach to coaddition, but will not delve into this issue extensively.

This topic has been discussed in the literature in many contexts, but unfortunately some of the key principles have not been explicitly demonstrated.  For example:
\begin{itemize}
\item \citet{2014ApJ...794..120A} provide some explanation without mathematical justification in section 3.4: ``To keep track of the variance of data images, pixel by
pixel inverse variance images are a natural choice, but
they produce biases in the resulting mean. At low signal to noise the upward fluctuations in signal are given
more weight than downward fluctuations as a result of
the one-sided nature of the Poisson distribution. This
bias is deterministic and one could correct for it, but,
for example, for $u$-band data with its 120$e^{-}$ of sky noise,
pixels at $1\sigma$ above sky would be biased by 0.5\% and this
would be a fair fraction of our photometric error budget. Another problem is that per-pixel inverse variance
weighting systematically changes the shape of the PSF
as a function of the magnitude of the object.''  This is another way of saying there is no well-defined PSF for the resulting image when using per-pixel inverse variance weighting.

\item \citet{2017ApJ...836..187Z} have a thorough discussion of the impact of coaddition choices on detection and measurements made on the coadd, but simply assert in their introduction that the variance used for weighting should be the ``variance of all the background noise sources (e.g., background, readout noise)''.  This is an implicit rejection of weighting that includes information about the astronomical sources such as stars and galaxies, but without explanation. This work discusses a coaddition scheme that satisfies that criterion and that was proposed by Nick Kaiser in an unpublished technical note \refresponse{\citep{Kaiser}}.

    \item \citet{2018ARA&A..56..393M} asserts without explanation in section 2.4 that depending on how coaddition is carried out, ``the coadd may not even have a well-defined PSF at each point (for example, inverse
variance weighting that depends on the total flux, or use of a median for the coadd). The
answer to this challenge is to generate the coadd in a principled way that results in a single
well-defined PSF at each point.'' This statement is consistent with the above two references, but is provided without explanation.

\item \citet{2018PASJ...70S...5B} asserts without illustration in section 3.3.2 that ``For PSF model coaddition to be valid, the operation used to
combine all input pixels at each point on the coadd image must
be strictly linear –{} robust estimators such as the median or
sigma-clipped means cannot be used. Nonlinear estimators do
not just prevent PSF coaddition from working, however; they
prevent the coadded image from even having a well-defined coadd PSF. Any estimator that rejects individual pixel outliers will tend to reject pixels in the cores of stars on the best-seeing exposures, and brighter stars will experience more rejection, giving
them a different profile than fainter stars. It should be emphasized that this occurs even in the absence of noise, and even
with extremely weak outlier rejection (e.g.\ clipping at $10\sigma$). All
robust estimators start from the ansatz that all input values are
drawn from the same underlying distribution, and convolution
with different PSFs means that they are not.''   This is an even more general statement, rejecting all nonlinear coaddition schemes; an explanation is given only for the problems with outlier rejection during coaddition, and not for the more general statements at the start of the quoted material. 

\item A \refresponse{Rubin Observatory} technical note about coaddition algorithms \refresponse{\citep{DMTN015}} states that when producing a coadd via a weighted summation over individual exposures, ``It is equally important that the weight function vary only on scales much larger than the size of the PSF; without this the coadd has no well-defined effective PSF.''

\item \citet{2011ApJ...741...46R} note in section 2.2 that they derive a linear coadd estimator because this ``ensures that the final PSF ... is independent of image brightness, so the stars that were not used in the PSF determination are tracers of how well the image synthesis worked.''

\end{itemize}

In this paper we aim to provide some of the mathematical justification and physical intuition behind these statements, beginning with defining formalism in Section~\ref{sec:formalism}, then discussing linear and nonlinear coaddition schemes in Sections~\ref{sec:lin} and~\ref{sec:nonlin}, respectively.

\section{Formalism}\label{sec:formalism}

The most basic definition of a ``point-spread function'' is that it is a function that describes the way that a bundle of rays, which would be incident on a single point, say $(x_0,y_0)$, spreads out spatially around that point. 
\irresponse{That is, if the incident profile is a Dirac delta function $\delta(x-x_0,y-y_0)$, then} the function $P(x,y)$ describes the resulting distribution of rays relative to the nominal point of incidence:
\begin{equation}
    \delta(x-x_0,y-y_0) \rightarrow P(x-x_0,y-y_0).
\end{equation}
\irresponse{Here the $(x,y)$ coordinates are not necessarily detector coordinates; they could instead be some idealized sky coordinate system after correction for astrometric distortions.}  
Across a typical astronomical image, this function $P$ may vary as a function of position $x_0,y_0$, but it typically does so fairly slowly.  Under the approximation that the functional form of the PSF is essentially unchanging across the extent of a single galaxy or other object or area of interest, the observed image can be described as a convolution:
\begin{align}
I(x,y) &= T(x,y) \otimes P(x,y),\\
&\equiv \int_{-\infty}^{\infty} \mathrm{d}x'
\int_{-\infty}^{\infty} \mathrm{d}y' T(x',y')P(x-x', y-y')
\end{align}
where $\otimes$ denotes convolution and $T(x,y)$ is the underlying true surface brightness distribution of the relevant sky scene.

\irresponse{\subsection{The Coadd PSF}}
In this work we assume that multiple background-subtracted images $I_i(x,y)$ (with subscript $i$ indexing the individual images) are being combined to produce a coadded image $I_\text{coadd}(x,y)$.   We also presume that each image has the same true underlying sky scene\footnote{This means that technically, our results only apply to static sources.  Following the results of this paper, it seems that the coadd PSF for time-varying sources will differ depending on the light curve (e.g., in the limit that a variable or transient source is very bright in just one exposure and faint in all the others, its coadd PSF will essentially be that of the exposure where it is very bright). \irresponse{This effect could even cause more basic issues for LSST.  The initial survey plan was that each visit (observation at a particular point) would consist of two 15-second ``snaps'' which would be combined into a single 30-second exposure. It is still to be decided whether this will be done, but if it is, then the issue of the PSF for time-varying objects in a coadd is} even relevant to how the two snaps are combined. Exploring these issues for time-varying sources is beyond the scope of this paper.} $T(x,y)$ (hence it has no index $i$) but a potentially different effective PSF (c.f.\ Sec~\ref{sec:sampling})
%\footnote{We use the phrase `effective PSF' to refer to the convolution of the PSF from the atmosphere and telescope optics with the pixel response function and with sensor effects that act as convolutions, such as charge diffusion.  The effective PSF is then sampled at the centers of pixels.  We prefer this convention over the alternative, not including the pixel response and then explicitly integrating over the pixel response function, as it makes the bookkeeping cleaner for coadded or resampled images. 
%Also note there is some ambiguity in the notation we are using for the $(x,y)$ argument of the effective PSF.  Here we mean that the effective PSF may vary as a function of position -- so when evaluating $P_i(x,y)$ at some specific $(x,y)$ position, we get a representation of the PSF that is also an image of some dimensionality.  (This is in contrast to $T(x,y)$, which when evaluated at a specific $x$ and $y$ is just a number.)} 
$P_i(x,y)$.  Finally, as mentioned above, we assume that the form of the PSF may vary on large spatial scales, but is assumed to not change \irresponse{significantly} across the extent of an individual star or galaxy or other local area of interest in the image.  Within a portion of the image small enough that these assumptions hold, the image amplitude prior to the addition of noise may be represented as
\begin{equation}
I_i(x,y) = T(x,y) \otimes P_i(x,y).
\end{equation}

\irresponse{We investigate the following question:} under what circumstances does the coaddition process \irresponse{of combining multiple images $I_i(x,y)$} produce a coadded image \irresponse{$I_\text{coadd}(x,y)$} such that there is a well defined function $P_\text{coadd}(x,y)$ that satisfies the condition
\begin{equation}\label{eq:coaddpsf}
    I_\text{coadd}(x,y) = T(x,y) \otimes P_\text{coadd}(x,y),
\end{equation}
or its Fourier-space equivalent 
\begin{equation}\label{eq:kcoaddpsf}
    \widetilde{I}_\text{coadd}(k_x,k_y) = \widetilde{T}(k_x,k_y) \widetilde{P}_\text{coadd}(k_x,k_y)?
\end{equation}
\irresponse{As we will see, the existence of a function $P_\text{coadd}(x,y)$ satisfying this condition puts quite strong requirements on the nature of the coaddition.  For instance, it implies that any stellar (point-source) images must scale linearly with the flux of the delta-function source.  This requirement is the origin of several tests we will carry out in Sections~\ref{sec:lin} and~\ref{sec:nonlin}. It also imposes requirements on extended sources, which we will explore further in Section~\ref{subsec:extended}.}

Each image is presumed to have a \refresponse{constant} background level $B_i$ that has been subtracted prior to PSF measurement and coaddition.  \refresponse{In a realistic case, the background is spatially-varying, and should be represented as $B_i(x,y)$, but it typically varies on larger scales than the astronomical objects of interest, and can be modeled well using empirical methods.}  This background level may potentially be used to define a \refresponse{weight function for coaddition, whether constant across each image or spatially varying}.  Moreover, while not included explicitly as a separate term, each image is presumed to have noise \irresponse{(the specific form of which is irrelevant to the rest of this discussion)}.

\subsection{Sampling}
\label{sec:sampling}
The above formalism treats the image as a continuous function of $(x,y)$.  In reality, one deals with pixellated images $I_i[m, n]$ where $m, n \in \mathbb{Z}$ denote the pixel index. A pixellated image is not \irresponse{merely the incident flux density} sampled at discrete $(x, y)$ locations, i.e., $I_i[m, n] \ne I_i(x=m, y=n)$. It is, rather, an integral of \irresponse{this function} within the pixel area. For simplicity, we illustrate this with a 1-dimensional \irresponse{flux density} $I_i(x)$, but the argument is applicable to 2-dimensional images as well.
For a uniform pixel spacing $a$, the pixel value at the $n^\text{th}$ pixel $I_i[n]$ is given by
\begin{equation}
    I_i[n] = \int_{(n-0.5)a}^{(n+0.5)a} \mathrm{d}x'' I_i(x'').
    \label{eq:pixellated_image}
\end{equation}
As a consequence, the total flux is conserved since
\begin{linenomath}\begin{equation}
    \sum_n I_i[n] = \int_{-\infty}^{\infty} dx'' I_i(x'').
\end{equation}\end{linenomath}

We now show that the pixellated image is still a convolution of the underlying scene with an effective PSF, sampled at the centers of pixels \citep[see also][]{Lauer99, Rowe15}. Starting from Eq.~\eqref{eq:pixellated_image},
\begin{linenomath}\begin{align*}
    I_i[n] &= \int_{(n-0.5)a}^{(n+0.5)a} \mathrm{d}x'' I_i(x'') \\
    &= \int_{(n-0.5)a}^{(n+0.5)a} \mathrm{d}x'' \int_{-\infty}^{\infty} \mathrm{d}x' T(x')P_i(x''-x').
\end{align*}\end{linenomath}
As before, $T$ is the underlying light profile of the \irresponse{sky scene}, and $P_i$ is the PSF in image $i$.  Next we explicitly introduce the pixel response function, $\Pi(u) = 1$ if $|u| < a/2$ and 0 otherwise:
\begin{linenomath}\begin{equation*}
    I_i[n]= \int_{-\infty}^{\infty} \mathrm{d}x' T(x') \left.\left[ \int_{-\infty}^{\infty} \mathrm{d}x'' P_i(x'' -x') \Pi(x-x'') \right|_{x=na}\right].
\end{equation*}\end{linenomath}
Finally, we change the variable of integration from $x''$ to $q=x''-x'$ in the inner integral,
\begin{linenomath}\begin{align}
    I_i[n]&= \int_{-\infty}^{\infty} \mathrm{d}x' T(x') \left.\left[ \int_{-\infty}^{\infty} \mathrm{d}q P_i(q) \Pi\left((x-x')-q\right) \right|_{x=na}\right] , \notag\\
    &= \int_{-\infty}^{\infty} \mathrm{d}x' T(x') P_{\text{eff},i}(x-x') \big|_{x=na},
\end{align}\end{linenomath}
\irresponse{where we define the effective PSF as the convolution of the PSF with the pixel response:}
\begin{equation}
    P_{\text{eff},i}(x) \equiv P_i(x) \otimes \Pi(x).
\end{equation}
\refresponse{For more about the distinction between the continuous flux distribution on the detector versus the sampled or effective PSF including pixelization, see, e.g., \cite{2000PASP..112.1360A}. }
We can thus consider the discrete pixel values as samples of the underlying \irresponse{scene} convolved with the effective PSF, \irresponse{evaluated at the pixel centers}.  However, for simplicity of notation, in the rest of this note we simply write $P(x,y)$ rather than $P_\text{eff}(x,y)$ to denote the effective PSF.

Since the convolution operation is associative, it is possible to include sensor effects such as charge diffusion, interpixel capacitance etc.\ as part of the effective PSF, as long as those effects act as convolutions (which is not the case for all sensor effects). We use the phrase ``effective PSF'' to refer to the convolution of the PSF from the atmosphere and telescope optics with the pixel response function and with sensor effects that act as convolutions.  We prefer this convention over the alternative, i.e. not including the pixel response and then explicitly integrating over the pixel response function, as it makes the bookkeeping cleaner for coadded or resampled images.

%If the PSF is Nyquist sampled, i.e., $\tilde{P}(k) \equiv 0 ~\forall~ |k| \ge \pi/a$, then so is the effective PSF. The Fourier transform of the pixel response has zeros at $k = m\pi/a$ for integer values of $m$. Since it is non-zero in the range where $\tilde{P}(k)$ is non-zero, it is possible to deconvolve and reverse\footnote{This is assuming effects that cannot be modeled as convolutions, and detector noise, are negligible.} the effect of the pixel response function.  The Nyquist theorem then guarantees that $I_i(x)$ can be recovered from the discrete samples $I_i[n]$. For the remainder of the paper, we therefore assume we can work with $I_i(x)$. The case where the PSF is not Nyquist sampled is beyond the scope of this paper.

%\mike{Proposed replacement for the above final para:}

Once we redefine $P(x,y)$ to really mean the effective PSF $P_\text{eff}(x,y)$, then $I_i(x,y)$ as defined in Eq.~\eqref{eq:coaddpsf} is a continuous function that, when sampled at the pixel centers, gives the observed image values.
Throughout the rest of the paper, we neglect this final step of sampling this function at the pixel centers, as it is irrelevant to the points we are making.  The results that we find apply to $I(x,y)$ will apply equally \irresponse{well} to the real $I[m,n]$ data, regardless of whether the PSF is Nyquist sampled or other similar details.

\section{Linear coaddition schemes}\label{sec:lin}

In this section, we presume that the images are summed with weight functions $w_i(x,y)$, such that
\begin{equation}\label{eq:lincoadd}
    I_\text{coadd}(x,y) = \frac{\sum_i w_i(x,y) I_i(x,y)}{\sum_i w_i(x,y)}.
\end{equation}
The denominator ensures flux conservation in the coaddition process by enforcing that the relative weights $w_i/\sum_i w_i$ always sum to 1. Note that this is not the most general linear coadd estimator; a matrix product would be more general, and in the case of undersampled images, a matrix product\footnote{Here the ``matrix product'' would connect pixels in the coadd with pixels in the input image.  For a list of $n_\text{in}$ input pixel values $\mathbf{d}_\text{in}$ in the set of images to be coadded, and a list of $n_\text{out}$ desired coadd pixel values $\mathbf{d}_\text{out}$, the most general relationship is $\mathbf{d}_\text{out} = \mathbf{W} \mathbf{d}_\text{in}$, where $\mathbf{W}$ is an $n_\text{out}\times n_\text{in}$ matrix.  In practice, even for approaches employing a matrix product, a variety of approximations are made to this general (but computationally unwieldy) formulation.} is actually necessary to get a well-defined coadd PSF \citep{2011ApJ...741...46R}.  However, for simplicity of notation we use Eq.~\eqref{eq:lincoadd} throughout this paper.

We begin with a direct illustration of the impact of weighting choices in a linear coaddition scheme, and then follow with a more formal demonstration of the circumstances under which the coadd has a well-defined PSF.

\subsection{Direct illustration}\label{subsec:direct}

\begin{figure*}
\begin{center}
\includegraphics[width=5.5in]{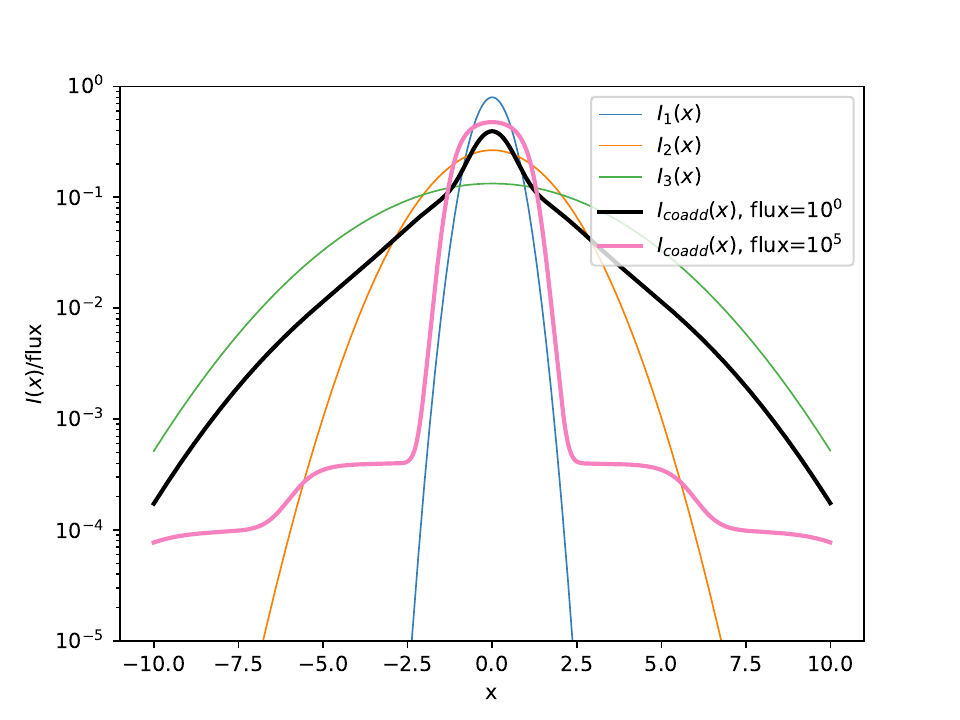}
 \caption{Result of coadding three one-dimensional images of stars with different PSFs using \irresponse{per-pixel} inverse variance weighting that \irresponse{depends on} the signal, for two different choices of the star flux $f_\text{star}$.  The figure shows the \irresponse{three input} 1D images \irresponse{in blue, orange, and green} for the three star images $I_1(x)$, $I_2(x)$, and $I_3(x)$ (Gaussians with $\sigma=0.5$, $1.5$ and $3.0$, respectively), all normalized to integrate to one for convenience of comparison. \irresponse{The resulting coadded images $I_\text{coadd}(x)$ for two different stellar fluxes are also shown, each} produced assuming a constant background level of $10$ in each image, with inverse variance weighting from Eq.~\eqref{eq:optweight} including both background and signal.  As shown, the \irresponse{normalized} coadded images differ depending on whether the images are background-dominated (lower star flux, shown in black) or signal-dominated at the star center (higher star flux, shown in pink).} \label{fig:coadd_psf}
\end{center}
 \end{figure*}
 
A natural choice for weighting is inverse variance weighting\footnote{Inverse variance weighting is {\em statistically} optimal \refresponse{for estimates of the flux in the case that the samples are drawn from the same underlying distribution,} unless one adopts more general linear coadd estimator involving a matrix product, as in the Kaiser approach \refresponse{\citep{Kaiser}}. 
 \refresponse{Given the assumption that the samples are drawn from the same underlying distribution, which is not generally the case for multiple observed images of the sky due to the differing PSF in those images, the inverse variance weight is not actually statistically optimal for coaddition in practice, besides the issues pointed out with this coaddition method in the main text.}}.  We therefore define a fully spatially \refresponse{varying weight} for each image:
%while ensuring the weights sum to 1 at each point:
\begin{equation}\label{eq:optweight}
    w_{i,\text{opt}}(x,y) = \frac{1}{B_i(x,y)+I_i(x,y)},
\end{equation}
\irresponse{where $B_i$ is the background level as before.} 
Combining Eqs.~\eqref{eq:lincoadd} and~\eqref{eq:optweight}, the coadded image with this \refresponse{inverse variance} weight is defined as
\begin{equation}\label{eq:optcoadd}
    I_\text{coadd,opt}(x,y) = \frac{\sum_i I_i(x,y)/\left[B_i(x,y)+I_i(x,y)\right]}{\sum_i 1/\left[B_i(x,y)+I_i(x,y)\right]}.
\end{equation}

To illustrate the impact of weight functions, we presume we are viewing an image of a single star.  In this case, we can assume that the sky scene consists of a \irresponse{Dirac} delta function at the center of our image, $(x_0,y_0) = (0,0)$, so  $T(x,y)=f_\text{star}\delta(x,y)$, and therefore $I_i(x,y)=f_\text{star}P_i(x,y)$ where $f_\text{star}$ is the total star flux.  We further assume that the PSFs and backgrounds in the images \irresponse{may} in general differ from each other, but are spatially constant, i.e., $B_i(x,y)=B_i$.  For simplicity, we consider three PSFs that are described as Gaussians with $\sigma=0.5$, $1.5$ and $3.0$ (arbitrary units), and using the same background level of $B=10$ for each.  We then ask the question of what does the coadd look like for different values of $f_\text{star}$?  The result is shown in Figure~\ref{fig:coadd_psf}.

As shown, the resulting coadded image differs depending on whether the star flux is such that the images are background-dominated at all positions, or whether there are some positions near the center of the star that are signal-dominated.  This is a key result: in reality, there are objects of different fluxes throughout the images, and this figure suggests that once the images have different PSFs, the dependence of the weights on the \irresponse{sources' fluxes} means that objects will appear differently in the coadds depending on their brightness, even in the very simple case that we only have \irresponse{three} images with the same constant background level and with a PSF that does not vary spatially within the images.  Once that is the case, there is no well-defined PSF for these images, as defined in Eq.~\eqref{eq:coaddpsf}.  Therefore, \refresponse{this weighting scheme is} fundamentally flawed from the standpoint of systematics; the lack of a well-defined coadd PSF will induce systematic biases in the characterization of object properties.

Ultimately the origin of these issues is that the weights that define the coadd in Eq.~\eqref{eq:optcoadd} depend on the input signal in the objects in the sky scene.  It seems possible that weights that depend only on quantities that are unrelated to the sky scene (example: background level and average seeing size) will not induce the same problems, which is the origin of the choices made in \citet{2014ApJ...794..120A} and \citet{2017ApJ...836..187Z}.  \irresponse{This statement that weighting by the background level should not cause problems with the coadd PSF is true under the assumption that the background level is dominated by other factors besides the astronomical objects (stars and galaxies) of interest, e.g., scattered light in the Earth's atmosphere or zodiacal light, which is generally the case.} By extension, a coadd with weights that depend on object signals will come closer to having a well-defined PSF in the limit that all objects are background-dominated.

 \subsection{Formal demonstration}\label{subsec:formal}

 A more principled mathematical view of the issue comes from the relationship between these quantities in Fourier space.  Starting from Eq.~\eqref{eq:lincoadd}, we can use
 \begin{linenomath}\begin{align*}
     I_\text{coadd}(x,y) &= \frac{\sum_i w_i(x,y) I_i(x,y)}{\sum_i w_i(x,y)} \\
     &= \frac{\sum_i w_i(x,y) \text{InvFT}[\widetilde{I}_i(k_x,k_y)]}{\sum_i w_i(x,y)}\\
     &= \frac{\sum_i w_i(x,y) \text{InvFT}[\widetilde{T}(k_x,k_y)\widetilde{P}_i(k_x,k_y)]}{\sum_i w_i(x,y)}.
 \end{align*}\end{linenomath}
And Eq.~\eqref{eq:coaddpsf} tells us that we want to know when is there some function $\widetilde{P}_\text{coadd}(k_x,k_y)$ such that
\begin{linenomath}\begin{equation*}
    I_\text{coadd}(x,y) = \text{InvFT}[\widetilde{T}(k_x,k_y)\widetilde{P}_\text{coadd}(k_x,k_y)]?
\end{equation*}\end{linenomath}
We could imagine trying to find $\widetilde{P}_\text{coadd}$ via division,
\begin{linenomath}\begin{equation}\label{eq:fourier_solve}
    \widetilde{P}_\text{coadd}(k_x,k_y) = \frac{\widetilde{I}_\text{coadd}(k_x,k_y)}{\widetilde{T}(k_x,k_y)}.
\end{equation}\end{linenomath}
However, for there to be a valid PSF, it needs to be the case that the \irresponse{(Fourier-transformed) true scene $\widetilde{T}$} divides out of this ratio. In other words, the ratio should be independent of $\widetilde{T}$.  It seems that for a completely general set of weights to result in a well-defined PSF for the coadd, $\widetilde{I}_\text{coadd}(k_x,k_y)$ must include the scene in the form of a straightforward product.  Or, equivalently, it must be possible to write $I_\text{coadd}(x,y)$ as the convolution of the true scene and some other function.   For a linear coaddition scheme, this will only occur under the following conditions:
\begin{itemize}
    \item \textbf{Same PSF in each image:} If all the component exposures have the same PSF (which could be approximately true, e.g., for space-based observations, where there is no atmosphere, if there is very little dithering or other effects such as thermal ``breathing'' between the exposures), then we have $P_i(x,y)=P(x,y)$.  In this case, Eq.~\eqref{eq:lincoadd} reduces to
    \begin{linenomath}\begin{align*}
         I_\text{coadd}(x,y) &= \frac{\sum_i w_i(x,y) T(x,y)\otimes P(x,y)}{\sum_i w_i(x,y)} \\
         &= T(x,y)\otimes P(x,y).
    \end{align*}\end{linenomath}
    Since each image has the same PSF, they are all the same (modulo noise).  Hence the sum over the spatially varying weight function in each image cancels out of the numerator and denominator, and has no impact on the PSF of the resulting coadd.  The coadded image has a well-defined PSF: it is the same as the PSF in each input image.
    
    \item \textbf{Spatially constant weight functions:}  If the weight for each exposure is not a function of position on the exposure, then  $w_i(x,y)=w_i$.  In this case, we can rewrite Eq.~\eqref{eq:lincoadd} as
    \begin{linenomath}\begin{align}
        I_\text{coadd}(x,y) &= \frac{\sum_i w_i I_i(x,y)}{\sum_i w_i} \nonumber\\
        &= \frac{\sum_i w_i \left[T(x,y) \otimes P_i(x,y)\right]}{\sum_i w_i} \nonumber \\
        &= T(x,y) \otimes \frac{\sum_i w_i P_i(x,y)}{\sum_i w_i},\label{eq:constweight}
    \end{align}\end{linenomath}
    where the last step relies on the linearity of convolution.  
    Eq.~\eqref{eq:constweight} shows that in this case, the scene is included in $I_\text{coadd}(x,y)$ as a direct convolution, which means the coadd has a well-defined PSF, 
    \begin{linenomath}\begin{equation}\label{eq:constweightpsf}
        P_\text{coadd}(x,y) = \frac{\sum_i w_i P_i(x,y)}{\sum_i w_i}.
    \end{equation}\end{linenomath}

    \irresponse{An example of a valid weighting scheme of this type would be if the weights were set to a spatially constant value determined based on the average PSF size in each contributing exposure.}
    
    \item \textbf{Weights that are independent of the true scene:}  If the weights are based solely on imaging characteristics (e.g., the background level, electronic read noise, etc.), and not anything correlated with the true scene, then the weights $w_i(x,y)$ are uncorrelated with $T(x,y)$.
    In this case, we have $\langle (w_i(x,y)-\bar w_i) T(x,y) \rangle = 0$, where \irresponse{the expectation value $\langle ...\rangle$ averages over the spatial extent of a large survey and} $\bar{w}_i = \langle w_i(x,y)\rangle$ is the spatial average of the weights.  
    
    Define $P_\text{coadd}$ to be the weighted average PSF:
    \begin{equation}\label{eq:varweightpsf}
        P_\text{coadd}(x,y) = \frac{\sum_i w_i(x,y) P_i(x,y)}{\sum_i w_i(x,y)}.
    \end{equation}
    It will not be the case now that we can satisfy Eq.~\eqref{eq:coaddpsf} specifically at every location, since $\widetilde{T}$ will not cancel out in Eq.~\eqref{eq:fourier_solve}, but we can satisfy it in an expectation-value sense, averaging over all locations in a survey (calculation shown in 1D rather than 2D for notational simplicity):
    \begin{linenomath}\begin{align*}
      \left\langle I_\text{coadd}(x) - T(x) \otimes P_\text{coadd}(x) \right\rangle
      &= \left\langle \frac{
                  \sum_i w_i(x) \int dx^\prime  T(x^\prime) P_i(x-x^\prime)  }{\sum_i w_i(x)} 
                  - \int\! dx^\prime  \frac{T(x^\prime) \sum_i w_i(x-x^\prime) P_i(x-x^\prime)}{\sum_i w_i(x-x^\prime)} \right\rangle \\
      &= \left\langle
                  \sum_i \int\! dx^\prime  T(x^\prime) P_i(x-x^\prime) \left(\frac{w_i(x)}{\sum_i w_i(x)} - \frac{w_i(x-x^\prime)}{\sum_i w_i(x-x^\prime)}\right) \right\rangle \\
      &=
                  \sum_i \int\! dx^\prime  T(x^\prime) P_i(x-x^\prime)
                  \left( \left\langle \frac{w_i(x)}{\sum_i w_i(x)}\right\rangle - \left\langle\frac{w_i(x-x^\prime)}{\sum_i w_i(x-x^\prime)}\right\rangle \right)\\
      &= 0
    \end{align*}\end{linenomath}
    where the second step (passing $T$ and $P_i$ out of the expectation brackets) relies on $w_i$ being uncorrelated with $T$ and $P_i$\footnote{\irresponse{Note that technically in addition to being uncorrelated with $T(x,y)$, the weights must also be uncorrelated with the PSF profile values $P_i(x,y)$, so $\langle (w_i(x,y)-\bar w_i) P_i(x,y) \rangle = 0$. For example, weight functions that depend on the spatial profile of the PSF around some particular central point would be disallowed. As this would be a bizarre choice for coaddition weights, we ignore this fairly pedantic technical point for the remainder of the paper. We note in particular that weights that depend on overall PSF characteristics such as size or shape are acceptable.}}.
    
    Thus, in this case, we have the slightly modified condition:
    \begin{linenomath}\begin{equation}\label{eq:expcoaddpsf}
        \left\langle I_\text{coadd}(x,y) \right\rangle = \left\langle T(x,y) \otimes P_\text{coadd}(x,y) \right\rangle
    \end{equation}\end{linenomath}
    Each location on the image will not specifically follow Eq.~\eqref{eq:coaddpsf}, but using this equation as a model of the observed coadd image will be unbiased when applied to a large ensemble of sources.  For \refresponse{many} purposes, this is a sufficient definition for there being a well-defined PSF.

\end{itemize}
These are the only options \irresponse{we were able to identify} that guarantee that the linear coadd defined in Eq.~\eqref{eq:lincoadd} can be rewritten as the convolution of the true sky scene with some other function. \irresponse{We do not believe that there are others where it is possible to} guarantee that the coadded image will have a well-defined PSF\footnote{\irresponse{Note that one could imagine devising spatially varying weights that depend on $T(x)$, yet cause the various terms in $\langle I_\text{coadd}(x)-T(x)\otimes P_\text{coadd}(x)\rangle$ to cancel for a specific choice of $T(x)$.  However, they would not be {\em guaranteed} to cancel in the real-world case that $T(x)$ is not perfectly known, unless the above-mentioned conditions apply.}}.
%\rachel{Unless I am missing something?} \mike{I couldn't think of any other cases that worked out either, but I don't immediately see a real proof of this claim.}

\irresponse{For real data,} it will never be the case that each image has precisely the same PSF; therefore, selecting spatially constant weight functions, or at least weight functions that are independent of the signal $T(x,y)$, provides a viable route to ensuring the coadd has a well-defined PSF.  
In practice, the last option seems to be the most common, since the various references in Section~\ref{sec:intro} of this paper accept the idea of weighting by the spatially-varying background, with the \irresponse{Rubin Data Management technical} note \refresponse{\citep{DMTN015}} providing an approximate criterion: the weight function should ``vary only on scales much larger than the size of the PSF.''  This criterion helps guarantee that the weights will be uncorrelated with the sources, at least for the majority of the astronomical objects in the images, with spatial extents that are typically just a few times the PSF size (for example, the small galaxies that are typically used for dark energy science).
%\rachel{We could explore this a little more with our formalism, with a weight function $w_i(x,y)$ to which we have applied a low-pass filter, and show what that does -- is this useful?}

%\irresponse{Our derivation above implies that one exception to the case pointed out in this technical note is that the weight function can vary on smaller scales than the PSF or galaxies if those variations are uncorrelated with the sky scene -- e.g., setting the weights to 0 in the locations of bad pixels.}

\section{Nonlinear coaddition schemes}\label{sec:nonlin}

There are several possible nonlinear coaddition schemes that one might consider adopting in reality.  While the discussion in Sec.~\ref{subsec:formal} makes it challenging to envision a nonlinear approach that would result in a well-defined PSF for the coadd, we discuss the specific issues with two potential nonlinear approaches below before explaining mathematically why it is impossible for a nonlinear coadd scheme to have a well-defined PSF.

\subsection{Median coadds}

\begin{figure}
\begin{center}
\includegraphics[width=5in]{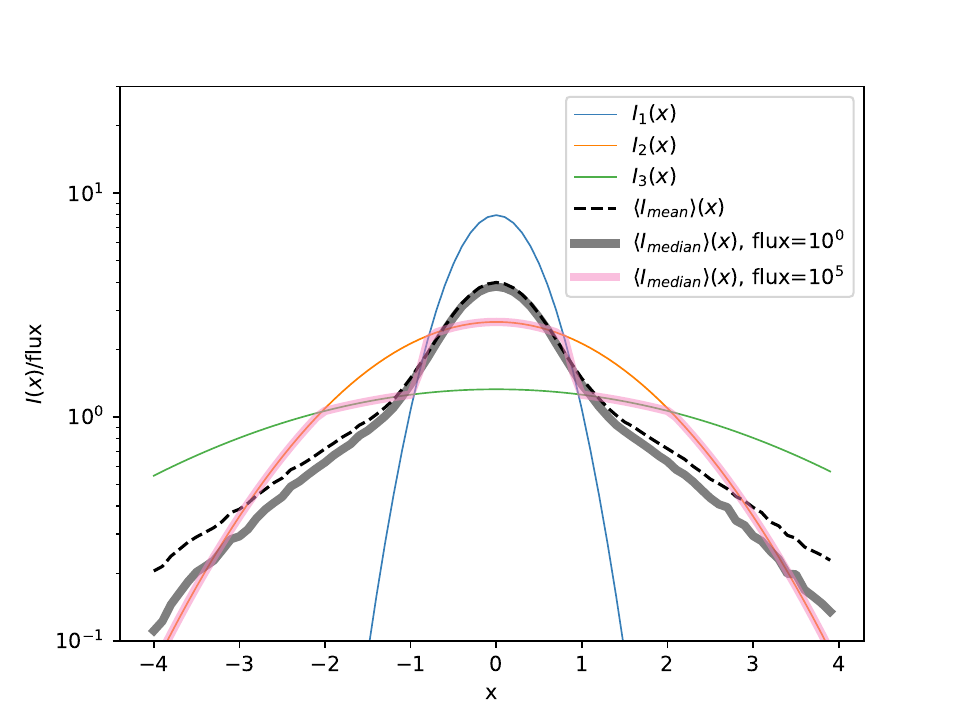}
 \caption{Result of applying a median coaddition procedure to three one-dimensional images of stars with different PSFs. 
 First, we show the images for three one-dimensional Gaussian images with different PSFs (denoted $I_i(x)$ in the legend) -- with the flux divided out to avoid stars of different fluxes requiring very different axes.  The other curves show average coadd profiles from many noisy realizations, including a sky background of 100.  The black dashed curve shows an unweighted mean coadd profile (for any star flux), which has a well-defined PSF.  The solid \irresponse{grey} curve shows the result for a median coadd for a low-flux (background-dominated) star, which as shown looks very similar to the mean coadd.  Finally, the pink curve shows the result for the median coadd for a high-flux (signal-dominated) star.  \irresponse{The median coadd curves are shown as very broad but semi-transparent lines in order to make it more clear that for various $x$ values they overlap various of the single-exposure images $I_i(x)$ or (for the low-flux case) the $\langle I_\text{mean}(x)\rangle$ curve.}
 The fact that the pink and \irresponse{grey} curves differ demonstrates that there is no well-defined PSF for a median coadd.
 }
 \label{fig:median_coadd_psf}
\end{center}
 \end{figure}

In order to avoid contamination by image artifacts, one possible ``robust statistic'' is a median rather than a mean.  In Figure~\ref{fig:median_coadd_psf} we illustrate the hazards of a median approach.
We show the images for three one-dimensional Gaussian images with different PSFs, along with the profile generated by a median coadd.  We generated one million noisy realizations of the images with a sky background of 100,  used them to make coadds, and then took the average of that coadd across all of those realizations to beat down the noise.  This was done for both low- and high-flux stars (flux = $1$ and $10^5$ respectively).
As shown, the high-flux case results in a star image with discontinuous slope, as expected from the medians of the three noise-free star images.  In contrast, in the low-flux case, the star image resembles an unweighted mean coadd. 
This illustrates that the PSF depends on the object magnitude, meaning that the coadd does not have a single well-defined PSF.

\begin{figure}
\begin{center}
    \includegraphics[width=5in]{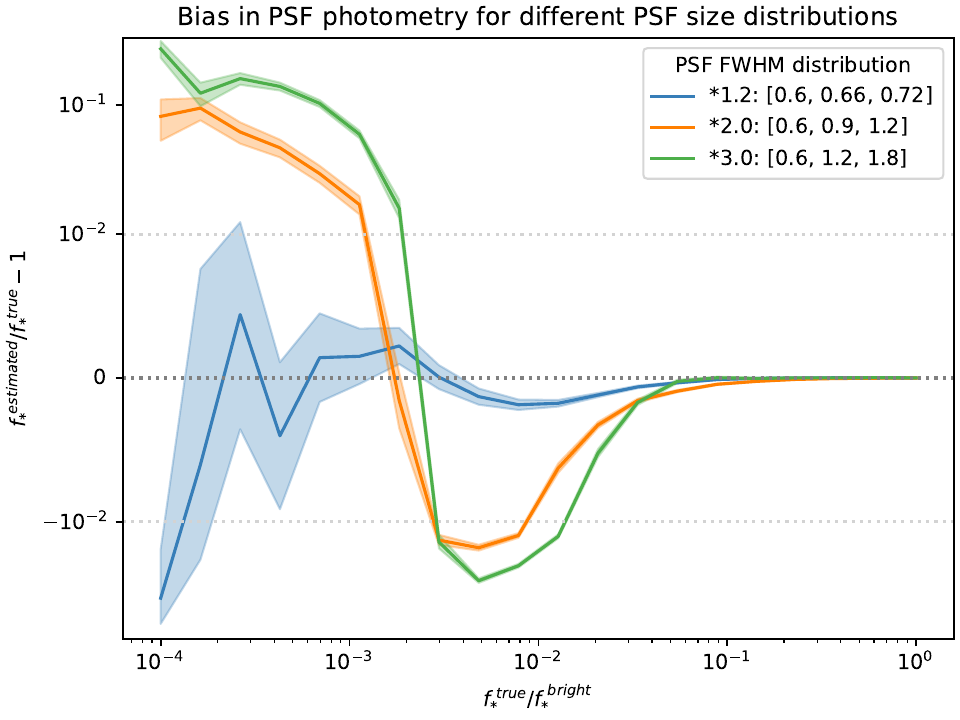}
    \caption{ Visualization of biases in PSF photometry for the median coaddition scheme, using 3 differing input exposures for the coadd.  Different colors indicate a different PSF size distribution. The legend shows a factor indicating the ratio of worst to best seeing size, along with the 3 PSF \irresponse{FWHM} values (\irresponse{Full-Width at Half-Maximum in arcsec, where the pixel scale is 0.2 arcsec}). An image of a star that is very bright compared to the background level is used as the PSF model, $P(x)$,  and fit to the image of a faint star to measure the PSF flux, as described in text. The curves show the average fractional bias in the estimated faint star flux, using 1000 independent noise realizations, as a function of the ratio of faint-to-bright star flux ratio.  The plot is shown on a symlog y-axis scale; light dotted grey lines indicate the range within which the y axis is linear, while the dark dotted grey line shows the ideal (bias-free) result of 0.  As expected, all curves converge to 0 as the faint star flux increases to match that of the bright star, but significant biases are present at lower flux, especially when the distribution of PSF sizes contributing to the coadd is relatively broad.}
    \label{fig:Median Coadd Bias}
\end{center}
\end{figure}

For a quantitative analysis of the impact of this coaddition scheme with more realistic inputs, Figure~\ref{fig:Median Coadd Bias} shows the bias introduced by the use of a median coadd when measuring point source photometry using PSF model flux measurements.
This algorithm uses a normalized bright star image as a PSF model, and finds the flux that best  describes the light profiles of other stars, minimizing the $\chi^2$ as a summation over image pixels with the same weight assigned to each pixel:
\begin{equation}
\chi^2 = \sum_\text{pixels} \frac{(I_\text{median}(x)-f_*P(x))^2}{\sigma^2},
\end{equation} 
where \irresponse{$I_\text{median}(x)$} is the image of the star for which we wish to measure the flux; $f_*$ is the fitting parameter, representing the apparent flux; $P(x)$ is the PSF model evaluated using a very bright star image; and  $\sigma$ is the background noise level in the image, which we take to be a constant.   Figure~\ref{fig:Median Coadd Bias} shows the fractional bias in these PSF flux estimates, as a function of the faint-to-bright star flux ratio, with means and uncertainties shown using fits to 1000 independent noise realizations.  
As expected, the bias is zero when the ``faint'' star flux approaches the bright star flux; but for fainter star fluxes,  biases in photometry appear, with a magnitude that depends on the distribution of seeing sizes in the images contributing to the coadd. When the seeing varies by only 20\% among the different images, biases in the PSF photometry are $\lesssim$1\%, comparable to the uncertainties. As the range of PSF sizes increases, so does the bias in PSF photometry, eventually reaching $\sim$10\% when the range is a factor of 2--3. Like other ground-based surveys, LSST is likely to have ratio around 2, for which the median coaddition scheme would cause unacceptable levels of bias in point-source photometry.

\subsection{Clipped-mean coadd}

Another obvious ``robust statistic'' to try is a clipped mean, i.e., defining some $\sigma$ value (through an \irresponse{inter-quartile range} or direct calculation), and omitting pixels from an image if they deviate from the mean by more than $N\sigma$, possibly working iteratively until convergence is achieved.
\irresponse{This is often a robust method for the problem of removing outliers from data with occasional glitches, e.g. cosmic rays in astronomical data.}

However, as noted in Sec.~\ref{sec:intro}, \citet{2018PASJ...70S...5B} has pointed out the issues with clipped means.  Functionally, they are the solution to a different problem, one where all of the data are drawn from the same underlying distribution.  In the case of coaddition, different images have different PSFs, so \irresponse{each} observed \irresponse{image $I_i(x,y)$} is drawn from a different distribution\irresponse{; therefore, a clipped mean is not actually a robust statistic for the problem of removing outlier data in the input images.}  For example, images with a smaller PSF will preferentially have higher pixel values at the centers of bright stars and galaxies, which will preferentially fail such an $N\sigma$ criterion orders of magnitude more frequently than one would expect for a Gaussian distribution in cases where the data are drawn from the same underlying distribution.  Omitting data preferentially as a function of PSF size, especially doing so differently for different parts of an object (i.e., core vs.\ wings), will naturally result in a coadd with no well-defined PSF. This is true even for relatively large values such as $N=10$.  

\subsection{Extended Sources}\label{subsec:extended}

The above arguments about median and clipped-mean coadds seem to imply that they are only a
problem at moderately low signal-to-noise.  At high signal-to-noise, the profiles of stars are nearly
independent of the flux of the star for both median and clipped-mean coadds.  
However, there is an additional problem with nonlinear coaddition schemes when considering extended sources (i.e., galaxies),
even when limiting to high signal-to-noise objects.

We typically determine the PSF profile by looking at the images of stars, since these represent
the response of the imaging process to a delta function.  This essentially acts like a
Green's function of the imaging ``operator''.
\begin{equation}
    P_i(x,y) = \mathbf{F}_i \delta(x,y),
\end{equation}
where $\mathbf{F}_i$ is an operator representing some potentially complicated imaging process.

We can then model the same process acting on an extended object
\begin{linenomath}\begin{equation}
    I_i(x,y) = \mathbf{F}_i T(x,y)
\end{equation}\end{linenomath}
as a convolution
\begin{linenomath}\begin{align}
    I_i(x,y) &= \mathbf{F}_i \int dx' dy' T(x^\prime,y^\prime) \delta(x-x^\prime,y-y^\prime) \nonumber\\
    &= \int dx' dy' T(x^\prime,y^\prime) \mathbf{F}_i \delta(x-x^\prime,y-y^\prime) \nonumber\\
    &= \int dx' dy'  T(x^\prime,y^\prime) P_i(x-x^\prime,y-y^\prime) \nonumber\\
    &= T(x,y) \otimes P_i(x,y),
\end{align}\end{linenomath}
which is valid so long as $\mathbf{F}_i$ is a linear operator.

When we coadd multiple images $I_i(x,y)$ into $I_\mathrm{coadd}(x,y)$, this coadding process
essentially becomes part of the imaging operator $\mathbf{F}_\mathrm{coadd}$:
\begin{linenomath}\begin{align}
    I_\mathrm{coadd}(x,y) &= \mathrm{Coadd} \left( \{ I_i(x,y) \} \right) \nonumber\\
    &= \mathrm{Coadd} \left( \{ \mathbf{F}_i(x,y) T(x,y) \} \right) \nonumber\\
    &\equiv \mathbf{F}_\mathrm{coadd} T(x,y).
\end{align}\end{linenomath}
When applied to stars, after normalizing the images by their fluxes,
we can call the resulting profile $P_\mathrm{coadd}$, but as we will see, this is
not really a PSF when $\mathbf{F}_\mathrm{coadd}$ is nonlinear.
\begin{linenomath}\begin{align}
    P_\mathrm{coadd}(x,y) &= \mathrm{Coadd} \left( \{ P_i(x,y) \} \right) \nonumber\\
    &= \mathrm{Coadd} \left( \{ \mathbf{F}_i(x,y) \delta(x,y) \} \right) \nonumber\\
    &= \mathbf{F}_\mathrm{coadd} \delta(x,y).
\end{align}\end{linenomath}
If the coadd process is not a linear combination of the inputs, 
then $\mathbf{F}_\mathrm{coadd}$ is not a linear operator, and it is no longer valid to
write $I_\mathrm{coadd}$ as a convolution of $P_\mathrm{coadd}$:
\begin{linenomath}\begin{align}
    I_\mathrm{coadd}(x,y) &= \mathbf{F}_\mathrm{coadd} \int dx' dy' T(x^\prime,y^\prime) \delta(x-x^\prime,y-y^\prime) \nonumber\\
    &\ne \int dx' dy' T(x^\prime,y^\prime) \left( \mathbf{F}_\mathrm{coadd} \delta(x-x^\prime,y-y^\prime) \right) \nonumber\\
    I_\mathrm{coadd}(x,y) &\ne T(x,y) \otimes P_\mathrm{coadd}(x,y).
\end{align}\end{linenomath}

\begin{figure}
\begin{center}
\includegraphics[width=5in]{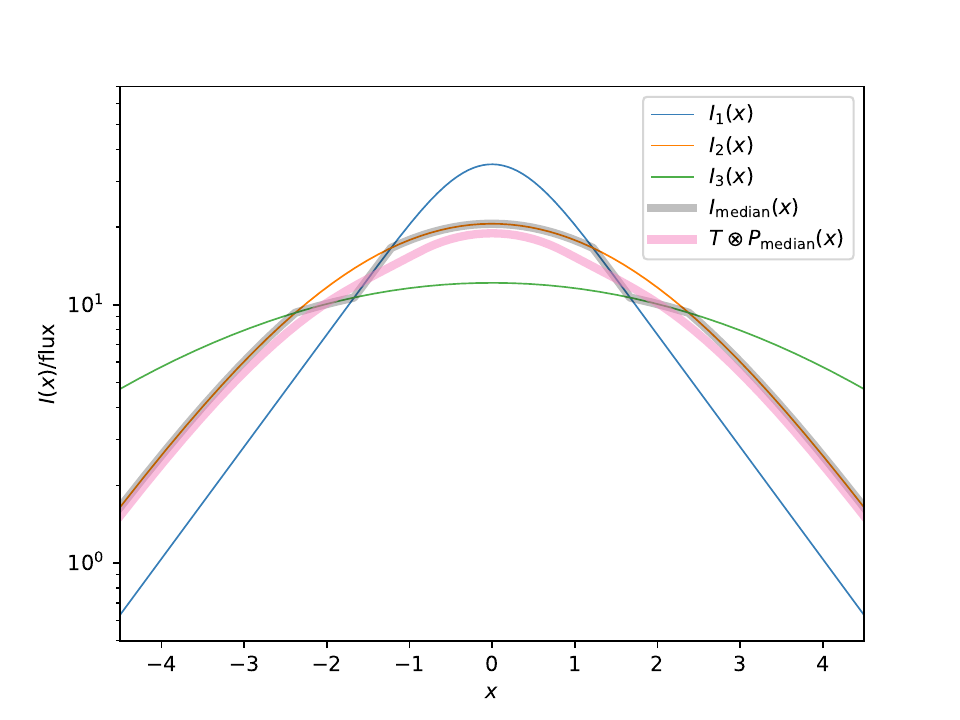}
 \caption{The effect of using a median coadd on an extended object. 
 The thin blue, orange, and green curves show a single exponential galaxy profile convolved by the 
 same three PSF profiles as shown in Figures~\ref{fig:coadd_psf} and \ref{fig:median_coadd_psf}. The \irresponse{broad, semi-transparent grey} curve is the result
 of using a median coadd procedure on these images.  The \irresponse{broad, semi-transparent} pink curve is the convolution of the true
 galaxy profile with the inferred PSF from the \irresponse{median} stellar profile at high signal-to-noise (i.e., the pink profile in Figure~\ref{fig:median_coadd_psf}).  The fact that the
 \irresponse{grey} and pink curves differ shows that even at high signal-to-noise, nonlinear coaddition algorithms
 (like median) do not result in a well-defined PSF.} \label{fig:extended}
\end{center}
 \end{figure}

Thus, when the coadd process is not linear,
the coadd image of an extended source is {\bf not} the convolution of the true scene by what
we would want to call the PSF.
Essentially, the entire concept of what we mean by a ``point-spread function'' (i.e., the response
of the imaging operator to a delta function, or ``point'') is invalid for a nonlinear imaging operator.  
Since the coaddition process is properly included as part of this operator, if that
step is nonlinear, then there is no well-defined PSF. 
Figure~\ref{fig:extended} demonstrates this fact for an exponential galaxy profile using a median
coaddition process (using the same three PSFs as shown in Figures~\ref{fig:coadd_psf} and \ref{fig:median_coadd_psf}).

\section{Conclusion}

In this paper, we showed a number of choices of coaddition algorithm that could result
in the coadd image not having a well-defined PSF, by which we mean that there is no
PSF function one can define for each location in the coadd such that the observed image is a convolution
of the true scene with this PSF. In general, for a coadd PSF to be well-defined, the coaddition procedure must be linear, and the weights assigned to each image in the linear coadd must be uncorrelated with the true sky scene.  For example, both of the following choices that are sometimes used in real data would result in a coadd that has a well-defined PSF: (a) a spatially constant weight that is inversely proportional to the average PSF size in an exposure, and (b) a spatially varying weight that is inversely proportional to the sky brightness ({\em not} including the astronomical sources).  

We note that not having a well-defined PSF is not necessarily a problem.  There are many use cases for coadd images
where one does not care whether the PSF is well-defined -- e.g., making color images
for showing observations of a planetary nebula, or possibly detection coadds where one may just want to maximize total signal-to-noise.
%\mike{??  I feel like at least some detection coadds might want a well-defined PSF, but maybe sometimes you don't really care?} \rachel{I think you are right, but since it says `possibly' and `may', I think it allows for the possibility that one may indeed want a well-defined PSF for a detection coadd in some cases.}
But for almost any use case where one will
be making some measurement (e.g., flux or shape) based on the pixel values in the coadd,
it is essential to have a well-defined PSF.  \refresponse{Examples of such applications include, but are not limited to, precise model-fitting estimates of galaxy photometry \citep[e.g.,][]{2018PASJ...70S...5B}; weak lensing shear measurements \citep[e.g.,][]{2008A&A...484...67P}; and joint modeling of quasars and the light profiles of their host galaxies \citep[see, e.g.,][]{2021ApJ...918...22L}.} In these cases, the tradeoff between systematic and statistical uncertainties may point towards choices that are less statistically optimal but that reduce systematic errors.  Note, moreover, that this has implications for steps that may often be considered part of a process that precedes coaddition, such as cosmic ray rejection (see, for example, the discussion in section 3.3 of \citealt{2018PASJ...70S...5B}).

We found that the following choices in the coadd algorithm lead to the coadd failing to
have a well-defined PSF:
\begin{itemize}
    \item All nonlinear coadd algorithms fail to have a well-defined PSF.  The coadd procedure is essentially part of the total imaging process, so if that is not linear in the inputs, then the final profiles of extended objects are not the convolution of the true scene with the profile of a delta function ("point"), thus invalidating the whole idea of a "point-spread function".
    \item Even for point sources, nonlinear coadd algorithms \irresponse{cause the expectation value of the observed profile shape to change} as a function of signal-to-noise, so the expected profile is not independent of flux.  Thus, they do not have a well-defined PSF even when only applied to stars.
    \item Linear coadd algorithms fail to have a well-defined PSF if their weights are a function of the signal.  This means if one is using inverse variance weighting, one should not include the Poisson noise of the signal as part of that variance.
    \item Technically, any weight function that is not spatially constant on each source image leads to the final image not having a formally well-defined PSF in the sense of following Eq.~\eqref{eq:coaddpsf}.  In practice this is \irresponse{often} not a problem in an expectation-value sense, so long as the spatial variability of the weights is uncorrelated with the input signal.  In this case, the coadd image follows Eq.~\eqref{eq:expcoaddpsf}. The non-uniform weights can add some additional scatter in the observed $I_\mathrm{coadd}(x,y)$, but not a bias. 
\end{itemize}

Given these constraints, we are left only with linear coadd schemes using weights that are independent of the signal.  If one needs the coadd PSF to be specifically correct at every location in the image, rather than merely unbiased, then one should use spatially invariant weights. \irresponse{A strategy that is usually acceptable in practice is to use weights that vary slowly across the image, so that the weights are nearly uniform for most objects of interest.  This may be particularly important for science involving two-point functions of measured quantities, since having a zero expectation value for the bias may not be sufficient to avoid biasing scientific analyses.} Note that we did not go into detail about matrix-based coadds, but these also qualify as linear, so they can be perfectly valid, so long as the matrix of weights follows the same restrictions \irresponse{as outlined above}.

\section*{Acknowledgments}

We thank Pat Burchat, Johann Cohen-Tanugi, Xiangchong Li, Josh Meyers, Morgan Schmitz, and Chris Stubbs for useful feedback on early  drafts of this paper. \refresponse{We also thank the anonymous journal referees, whose feedback resulted in improvements to the paper prior to publication.} The scripts that produce the plots in this paper are available at \url{https://github.com/LSSTDESC/DESCNote\_CoaddPsf}.  This paper has undergone internal review in the LSST Dark Energy Science Collaboration. The internal reviewers were \irresponse{Gary Bernstein, Johann Cohen-Tanugi, and Ismael Mendoza.}

RM and TZ acknowledge the support of the Department of Energy grant DE-SC0010118.
MJ is partially supported by
Department of Energy grant DE-SC0007901.

Author contributions to this work are as follows: RM synthesized telecon discussions and material from the literature to produce initial draft Note, discussed with co-authors, edited.  MJ wrote significant portions of the text, including 2.1, part of 3.2, 4.3, figure 4, plus edits and word smithing throughout.  RHL contributed to the study which led to this paper, and provided comments/suggestions during the writing of the paper. JB did some of the early work (on the HSC pipelines) that first identified the problems more fully described in this paper, and participated in discussions and editing of early drafts.  AK contributed to some of clarification and general feedback to the publication, and created a notebook to numerically verify some of the key equations (Eqs.\ 10-11). MDM created the median coadd bias quantification in Figure 3, as well as supporting text in Section 4.1. TZ double-checked the nonlinear weighting coadd with more specifications, and provided feedback on the impact of median coadd results.

The DESC acknowledges ongoing support from the Institut National de 
Physique Nucl\'eaire et de Physique des Particules in France; the 
Science \& Technology Facilities Council in the United Kingdom; and the
Department of Energy, the National Science Foundation, and the LSST 
Corporation in the United States.  DESC uses resources of the IN2P3 
Computing Center (CC-IN2P3--Lyon/Villeurbanne - France) funded by the 
Centre National de la Recherche Scientifique; the National Energy 
Research Scientific Computing Center, a DOE Office of Science User 
Facility supported by the Office of Science of the U.S.\ Department of
Energy under Contract No.\ DE-AC02-05CH11231; STFC DiRAC HPC Facilities, 
funded by UK BEIS National E-infrastructure capital grants; and the UK 
particle physics grid, supported by the GridPP Collaboration.  This 
work was performed in part under DOE Contract DE-AC02-76SF00515.

\section*{Author Affiliations}

\noindent $^1$McWilliams Center for Cosmology, Department of Physics, Carnegie Mellon University, Pittsburgh, PA 15213, USA \\
$^2$Department of Physics \& Astronomy, University of Pennsylvania, 209 South 33rd Street, Philadelphia, PA 19104-6396, USA\\
$^3$Department of Astrophysical Sciences, Princeton University, 4 Ivy Lane, Princeton, NJ 08544, USA\\

\bibliographystyle{style_and_logos/apj}
\bibliography{main}  

\begin{thebibliography}{}
\expandafter\ifx\csname natexlab\endcsname\relax\def\natexlab#1{#1}\fi

\bibitem[{{Anderson} \& {King}(2000)}]{2000PASP..112.1360A}
{Anderson}, J., \& {King}, I.~R. 2000, \pasp, 112, 1360

\bibitem[{{Annis} {et~al.}(2014){Annis}, {Soares-Santos}, {Strauss}, {Becker},
  {Dodelson}, {Fan}, {Gunn}, {Hao}, {Ivezi{\'c}}, {Jester}, {Jiang},
  {Johnston}, {Kubo}, {Lampeitl}, {Lin}, {Lupton}, {Miknaitis}, {Seo}, {Simet},
  \& {Yanny}}]{2014ApJ...794..120A}
{Annis}, J., {Soares-Santos}, M., {Strauss}, M.~A., {et~al.} 2014, \apj, 794,
  120

\bibitem[{Bosch(2016)}]{DMTN015}
Bosch, J. 2016, \url{https://dmtn-015.lsst.io/}

\bibitem[{{Bosch} {et~al.}(2018){Bosch}, {Armstrong}, {Bickerton}, {Furusawa},
  {Ikeda}, {Koike}, {Lupton}, {Mineo}, {Price}, {Takata}, {Tanaka}, {Yasuda},
  {AlSayyad}, {Becker}, {Coulton}, {Coupon}, {Garmilla}, {Huang}, {Krughoff},
  {Lang}, {Leauthaud}, {Lim}, {Lust}, {MacArthur}, {Mandelbaum}, {Miyatake},
  {Miyazaki}, {Murata}, {More}, {Okura}, {Owen}, {Swinbank}, {Strauss},
  {Yamada}, \& {Yamanoi}}]{2018PASJ...70S...5B}
{Bosch}, J., {Armstrong}, R., {Bickerton}, S., {et~al.} 2018, \pasj, 70, S5

\bibitem[{{Ivezi{\'c}} {et~al.}(2019){Ivezi{\'c}}, {Kahn}, {Tyson}, {Abel},
  {Acosta}, {Allsman}, {Alonso}, {AlSayyad}, {Anderson}, {Andrew}, {Angel},
  {Angeli}, {Ansari}, {Antilogus}, {Araujo}, {Armstrong}, {Arndt}, {Astier},
  {Aubourg}, {Auza}, {Axelrod}, {Bard}, {Barr}, {Barrau}, {Bartlett}, {Bauer},
  {Bauman}, {Baumont}, {Bechtol}, {Bechtol}, {Becker}, {Becla}, {Beldica},
  {Bellavia}, {Bianco}, {Biswas}, {Blanc}, {Blazek}, {Blandford}, {Bloom},
  {Bogart}, {Bond}, {Booth}, {Borgland}, {Borne}, {Bosch}, {Boutigny},
  {Brackett}, {Bradshaw}, {Brandt}, {Brown}, {Bullock}, {Burchat}, {Burke},
  {Cagnoli}, {Calabrese}, {Callahan}, {Callen}, {Carlin}, {Carlson},
  {Chandrasekharan}, {Charles-Emerson}, {Chesley}, {Cheu}, {Chiang}, {Chiang},
  {Chirino}, {Chow}, {Ciardi}, {Claver}, {Cohen-Tanugi}, {Cockrum}, {Coles},
  {Connolly}, {Cook}, {Cooray}, {Covey}, {Cribbs}, {Cui}, {Cutri}, {Daly},
  {Daniel}, {Daruich}, {Daubard}, {Daues}, {Dawson}, {Delgado}, {Dellapenna},
  {de Peyster}, {de Val-Borro}, {Digel}, {Doherty}, {Dubois},
  {Dubois-Felsmann}, {Durech}, {Economou}, {Eifler}, {Eracleous}, {Emmons},
  {Fausti Neto}, {Ferguson}, {Figueroa}, {Fisher-Levine}, {Focke}, {Foss},
  {Frank}, {Freemon}, {Gangler}, {Gawiser}, {Geary}, {Gee}, {Geha}, {Gessner},
  {Gibson}, {Gilmore}, {Glanzman}, {Glick}, {Goldina}, {Goldstein}, {Goodenow},
  {Graham}, {Gressler}, {Gris}, {Guy}, {Guyonnet}, {Haller}, {Harris},
  {Hascall}, {Haupt}, {Hernandez}, {Herrmann}, {Hileman}, {Hoblitt}, {Hodgson},
  {Hogan}, {Howard}, {Huang}, {Huffer}, {Ingraham}, {Innes}, {Jacoby}, {Jain},
  {Jammes}, {Jee}, {Jenness}, {Jernigan}, {Jevremovi{\'c}}, {Johns}, {Johnson},
  {Johnson}, {Jones}, {Juramy-Gilles}, {Juri{\'c}}, {Kalirai}, {Kallivayalil},
  {Kalmbach}, {Kantor}, {Karst}, {Kasliwal}, {Kelly}, {Kessler}, {Kinnison},
  {Kirkby}, {Knox}, {Kotov}, {Krabbendam}, {Krughoff}, {Kub{\'a}nek},
  {Kuczewski}, {Kulkarni}, {Ku}, {Kurita}, {Lage}, {Lambert}, {Lange},
  {Langton}, {Le Guillou}, {Levine}, {Liang}, {Lim}, {Lintott}, {Long},
  {Lopez}, {Lotz}, {Lupton}, {Lust}, {MacArthur}, {Mahabal}, {Mandelbaum},
  {Markiewicz}, {Marsh}, {Marshall}, {Marshall}, {May}, {McKercher}, {McQueen},
  {Meyers}, {Migliore}, {Miller}, {Mills}, {Miraval}, {Moeyens}, {Moolekamp},
  {Monet}, {Moniez}, {Monkewitz}, {Montgomery}, {Morrison}, {Mueller},
  {Muller}, {Mu{\~n}oz Arancibia}, {Neill}, {Newbry}, {Nief}, {Nomerotski},
  {Nordby}, {O'Connor}, {Oliver}, {Olivier}, {Olsen}, {O'Mullane}, {Ortiz},
  {Osier}, {Owen}, {Pain}, {Palecek}, {Parejko}, {Parsons}, {Pease},
  {Peterson}, {Peterson}, {Petravick}, {Libby Petrick}, {Petry},
  {Pierfederici}, {Pietrowicz}, {Pike}, {Pinto}, {Plante}, {Plate}, {Plutchak},
  {Price}, {Prouza}, {Radeka}, {Rajagopal}, {Rasmussen}, {Regnault}, {Reil},
  {Reiss}, {Reuter}, {Ridgway}, {Riot}, {Ritz}, {Robinson}, {Roby}, {Roodman},
  {Rosing}, {Roucelle}, {Rumore}, {Russo}, {Saha}, {Sassolas}, {Schalk},
  {Schellart}, {Schindler}, {Schmidt}, {Schneider}, {Schneider}, {Schoening},
  {Schumacher}, {Schwamb}, {Sebag}, {Selvy}, {Sembroski}, {Seppala}, {Serio},
  {Serrano}, {Shaw}, {Shipsey}, {Sick}, {Silvestri}, {Slater}, {Smith},
  {Smith}, {Sobhani}, {Soldahl}, {Storrie-Lombardi}, {Stover}, {Strauss},
  {Street}, {Stubbs}, {Sullivan}, {Sweeney}, {Swinbank}, {Szalay}, {Takacs},
  {Tether}, {Thaler}, {Thayer}, {Thomas}, {Thornton}, {Thukral}, {Tice},
  {Trilling}, {Turri}, {Van Berg}, {Vanden Berk}, {Vetter}, {Virieux},
  {Vucina}, {Wahl}, {Walkowicz}, {Walsh}, {Walter}, {Wang}, {Wang}, {Warner},
  {Wiecha}, {Willman}, {Winters}, {Wittman}, {Wolff}, {Wood-Vasey}, {Wu},
  {Xin}, {Yoachim}, \& {Zhan}}]{2019ApJ...873..111I}
{Ivezi{\'c}}, {\v{Z}}., {Kahn}, S.~M., {Tyson}, J.~A., {et~al.} 2019, \apj,
  873, 111

\bibitem[{Kaiser(2001)}]{Kaiser}
Kaiser, N. 2001,
  \url{http://pan-starrs.ifa.hawaii.edu/project/people/kaiser/imageprocessing/im++.pdf}

\bibitem[{Lauer(1999)}]{Lauer99}
Lauer, T.~R. 1999, Publications of the Astronomical Society of the Pacific,
  111, 227 – 237, cited by: 82

\bibitem[{{Li} {et~al.}(2021){Li}, {Silverman}, {Ding}, {Strauss}, {Goulding},
  {Birrer}, {Yesuf}, {Xue}, {Kawinwanichakij}, {Matsuoka}, {Toba}, {Nagao},
  {Schramm}, \& {Inayoshi}}]{2021ApJ...918...22L}
{Li}, J., {Silverman}, J.~D., {Ding}, X., {et~al.} 2021, \apj, 918, 22

\bibitem[{{LSST Science Collaboration} {et~al.}(2009){LSST Science
  Collaboration}, {Abell}, {Allison}, {Anderson}, {Andrew}, {Angel}, {Armus},
  {Arnett}, {Asztalos}, {Axelrod}, {Bailey}, {Ballantyne}, {Bankert},
  {Barkhouse}, {Barr}, {Barrientos}, {Barth}, {Bartlett}, {Becker}, {Becla},
  {Beers}, {Bernstein}, {Biswas}, {Blanton}, {Bloom}, {Bochanski}, {Boeshaar},
  {Borne}, {Bradac}, {Brandt}, {Bridge}, {Brown}, {Brunner}, {Bullock},
  {Burgasser}, {Burge}, {Burke}, {Cargile}, {Chandrasekharan}, {Chartas},
  {Chesley}, {Chu}, {Cinabro}, {Claire}, {Claver}, {Clowe}, {Connolly}, {Cook},
  {Cooke}, {Cooray}, {Covey}, {Culliton}, {de Jong}, {de Vries}, {Debattista},
  {Delgado}, {Dell'Antonio}, {Dhital}, {Di Stefano}, {Dickinson}, {Dilday},
  {Djorgovski}, {Dobler}, {Donalek}, {Dubois-Felsmann}, {Durech},
  {Eliasdottir}, {Eracleous}, {Eyer}, {Falco}, {Fan}, {Fassnacht}, {Ferguson},
  {Fernandez}, {Fields}, {Finkbeiner}, {Figueroa}, {Fox}, {Francke}, {Frank},
  {Frieman}, {Fromenteau}, {Furqan}, {Galaz}, {Gal-Yam}, {Garnavich},
  {Gawiser}, {Geary}, {Gee}, {Gibson}, {Gilmore}, {Grace}, {Green}, {Gressler},
  {Grillmair}, {Habib}, {Haggerty}, {Hamuy}, {Harris}, {Hawley}, {Heavens},
  {Hebb}, {Henry}, {Hileman}, {Hilton}, {Hoadley}, {Holberg}, {Holman},
  {Howell}, {Infante}, {Ivezic}, {Jacoby}, {Jain}, {R}, {Jedicke}, {Jee},
  {Garrett Jernigan}, {Jha}, {Johnston}, {Jones}, {Juric}, {Kaasalainen},
  {Styliani}, {Kafka}, {Kahn}, {Kaib}, {Kalirai}, {Kantor}, {Kasliwal},
  {Keeton}, {Kessler}, {Knezevic}, {Kowalski}, {Krabbendam}, {Krughoff},
  {Kulkarni}, {Kuhlman}, {Lacy}, {Lepine}, {Liang}, {Lien}, {Lira}, {Long},
  {Lorenz}, {Lotz}, {Lupton}, {Lutz}, {Macri}, {Mahabal}, {Mandelbaum},
  {Marshall}, {May}, {McGehee}, {Meadows}, {Meert}, {Milani}, {Miller},
  {Miller}, {Mills}, {Minniti}, {Monet}, {Mukadam}, {Nakar}, {Neill}, {Newman},
  {Nikolaev}, {Nordby}, {O'Connor}, {Oguri}, {Oliver}, {Olivier}, {Olsen},
  {Olsen}, {Olszewski}, {Oluseyi}, {Padilla}, {Parker}, {Pepper}, {Peterson},
  {Petry}, {Pinto}, {Pizagno}, {Popescu}, {Prsa}, {Radcka}, {Raddick},
  {Rasmussen}, {Rau}, {Rho}, {Rhoads}, {Richards}, {Ridgway}, {Robertson},
  {Roskar}, {Saha}, {Sarajedini}, {Scannapieco}, {Schalk}, {Schindler},
  {Schmidt}, {Schmidt}, {Schneider}, {Schumacher}, {Scranton}, {Sebag},
  {Seppala}, {Shemmer}, {Simon}, {Sivertz}, {Smith}, {Allyn Smith}, {Smith},
  {Spitz}, {Stanford}, {Stassun}, {Strader}, {Strauss}, {Stubbs}, {Sweeney},
  {Szalay}, {Szkody}, {Takada}, {Thorman}, {Trilling}, {Trimble}, {Tyson}, {Van
  Berg}, {Vanden Berk}, {VanderPlas}, {Verde}, {Vrsnak}, {Walkowicz},
  {Wandelt}, {Wang}, {Wang}, {Warner}, {Wechsler}, {West}, {Wiecha},
  {Williams}, {Willman}, {Wittman}, {Wolff}, {Wood-Vasey}, {Wozniak}, {Young},
  {Zentner}, \& {Zhan}}]{2009arXiv0912.0201L}
{LSST Science Collaboration}, {Abell}, P.~A., {Allison}, J., {et~al.} 2009,
  arXiv e-prints, arXiv:0912.0201

\bibitem[{{Mandelbaum}(2018)}]{2018ARA&A..56..393M}
{Mandelbaum}, R. 2018, \araa, 56, 393

\bibitem[{{Paulin-Henriksson} {et~al.}(2008){Paulin-Henriksson}, {Amara},
  {Voigt}, {Refregier}, \& {Bridle}}]{2008A&A...484...67P}
{Paulin-Henriksson}, S., {Amara}, A., {Voigt}, L., {Refregier}, A., \&
  {Bridle}, S.~L. 2008, \aap, 484, 67

\bibitem[{{Rowe} {et~al.}(2011){Rowe}, {Hirata}, \&
  {Rhodes}}]{2011ApJ...741...46R}
{Rowe}, B., {Hirata}, C., \& {Rhodes}, J. 2011, \apj, 741, 46

\bibitem[{{Rowe} {et~al.}(2015){Rowe}, {Jarvis}, {Mandelbaum}, {Bernstein},
  {Bosch}, {Simet}, {Meyers}, {Kacprzak}, {Nakajima}, {Zuntz}, {Miyatake},
  {Dietrich}, {Armstrong}, {Melchior}, \& {Gill}}]{Rowe15}
{Rowe}, B.~T.~P., {Jarvis}, M., {Mandelbaum}, R., {et~al.} 2015, Astronomy and
  Computing, 10, 121

\bibitem[{{Zackay} \& {Ofek}(2017{\natexlab{a}})}]{2017ApJ...836..187Z}
{Zackay}, B., \& {Ofek}, E.~O. 2017{\natexlab{a}}, \apj, 836, 187

\bibitem[{{Zackay} \& {Ofek}(2017{\natexlab{b}})}]{2017ApJ...836..188Z}
---. 2017{\natexlab{b}}, \apj, 836, 188

\end{thebibliography}

\end{document}